\documentclass[aps,prb,twocolumn,groupedaddress]{revtex4}
\usepackage{graphicx}
\begin{document}

\title{Universal scaling behavior of the c-axis resistivity of high-temperature superconductors}

\author{Y. H. Su}
\affiliation{Center for Advanced Studies, Tsinghua University,
Beijing 100084, China}
\author{H. G. Luo}
\affiliation{Institute of Theoretical Physics and
Interdisciplinary Centre of Theoretical Studies, Chinese Academy
of Sciences, P.O. Box 2735, Beijing 100080, China}
\author{T. Xiang}
\affiliation{Institute of Theoretical Physics and
Interdisciplinary Centre of Theoretical Studies, Chinese Academy
of Sciences, P.O. Box 2735, Beijing 100080, China}
\affiliation{Center for Advanced Studies, Tsinghua University,
Beijing 100084, China}


\begin{abstract}
We propose and show that the c-axis transport in high-temperature
superconductors is controlled by the pseudogap energy and the
c-axis resistivity satisfies a universal scaling law in the
pseudogap phase. We derived approximately a scaling function for
the c-axis resistivity and found that it fits well with the
experimental data of Bi$_2$Sr$_2$CaCu$_2$O$_{8+\delta}$,
Bi$_2$Sr$_2$Ca$_2$Cu$_3$O$_{10+\delta}$, and
YBa$_2$Cu$_3$O$_{7-\delta}$. Our works reveals the physical origin
of the semiconductor-like behavior of the c-axis resistivity and
suggests that the c-axis hopping is predominantly coherent.
\end{abstract}


\maketitle

\section{Introduction}
In the pseudogap phase of high-temperature (high-T$_c$)
superconductors, the temperature dependence of the c-axis
resistivity $\rho_c$ is semiconductor-like ($d\rho_c/dT < 0$), in
contrast to the in-plane resistivity $\rho_{ab}$ whose temperature
dependence is metal-like ($d\rho_c/dT > 0$). This dramatic
difference between $\rho_c$ and $\rho_{ab}$ is not what one might
expect within the conventional Fermi liquid theory. It has
stimulated vast theoretical and experimental investigations on the
interlayer dynamics of high-T$_c$ cuprates. However, the existing
theoretical models based on the notion of dynamic confinement
\cite{anderson97,kumar92} or incoherent interlayer hopping
\cite{leggett01} could not give a natural and unified explanation
for experimental data.

In high-T$_c$ cuprates, the c-axis hopping integral is highly
anisotropic and depends strongly on the in-plane momenta.
\cite{andersen95, chak93, xiang96, xiang98} Based on electron
structure calculations \cite{andersen95} and symmetry analysis,
\cite{xiang96, xiang98} it was shown that the c-axis hopping
integral in high-T$_c$ cuprates with tetragonal symmetry is given
by
\begin{equation}
t_{c}(k) \propto (\cos k_x - \cos k_y)^2, \label{tc}
\end{equation}
This peculiar in-plane momentum dependence of $t_c(k)$ results
from the hybridization between the bonding O 2p orbitals and Cu 4s
or $3d_{3z^2 - r^2}$ orbitals in each CuO$_2$ plane. It was
confirmed by the angle-resolved photoemission spectroscopy (ARPES)
measurements. \cite{feng01} $t_c$ vanishes along the two diagonals
of the Brillouin zone. These are also the directions where the
nodes of the $d_{x^2 - y^2}$ (d)-wave superconducting gap or the
normal state pseudogap (it has also the d-wave symmetry
\cite{loeser96, ding96}) are located. Thus the quasiparticle
excitations around the gap nodes have almost no contributions to
the c-axis hopping. This indicates that the c-axis dynamics is
governed by the quasiparticle excitations around the antinodal
points.

The interplay between $t_c(k)$ and the d-wave energy gap can
affect strongly the c-axis dynamics of electrons. In the
superconducting state, it leads to, for example, a T$^5$
temperature dependent c-axis superfluid density in low
temperatures, \cite{xiang96} in contrast to the linear-T behavior
of the in-plane superfluid density. In a pseudogap normal state,
since the contribution from the quasiparticle excitations around
the nodal points is suppressed by the integral (1), the interlayer
transport would behave as in a gapped system and $\rho_c$ should
show a thermally activated behavior. On the contrary, the in-plane
dynamics is governed by low-lying excitations around the gap nodes
in low temperatures. Thus $\rho_{ab}$ would behave similarly as in
a gapless system. This explains naturally why the temperature
dependence of $\rho_{c}$ is semiconductor-like while $\rho_{ab}$
is metal-like in the pseudogap phase. In making this argument, we
have implicitly assumed that the c-axis hopping is predominantly
coherent. \cite{alexandrov96} We believe this assumption is
correct, at least in the limit of weak impurity scattering. It is
supported by the observation of the bilayer splitting of the
bonding and antibonding bands in the ARPES spectra, \cite{feng01}
the intrinsic T$^5$ temperature dependence of the c-axis
superfluid density, as well as the recent angular-dependent
magnetoresistance measurement. \cite{hussey03}

The c-axis dynamics could be also affected by phonon or other
collective excitations. Along the nodal directions, a dispersion
kink has been observed in the spectral function of electrons by
the ARPES in both superconducting \cite{bogdanov00} and normal
states. \cite{lanzara01} This kink is believed to result from the
electron-phonon coupling. Along the antinodal direction, a
stronger kink feature was observed in the single-electron spectrum
below T$_c$. \cite{gromko03} This stronger kink seems to be
correlated with the peak-dip-hump lineshape as observed in the
ARPES spectra at antinodal points. \cite{camp99} It is likely to
be due to the coupling of electrons with a magnetic resonance
mode. It could also be explained by the coupling of electrons with
a B$_{1g}$-phonon mode. \cite{cuk04} However, this dispersion kink
together with the peak-dip-hump structure in the ARPES spectra
disappears above T$_c$. It suggests that the coupling between
electrons and phonons or magnetic resonance modes is strong around
the gap nodes but very weak along the antinodal directions in the
normal state. Thus we believe that the coupling between electrons
and phonons or magnetic resonance modes should have very weak or
negligible effect on c-axis transport properties in the normal
state.

\section{Scaling Hypothesis}
The above discussion indicates that in the normal state the
pseudogap is the only energy scale governing the c-axis hopping
around the antinodal regions. If the c-axis hopping is
predominately coherent, this would suggest that the temperature
dependence of $\rho_c$ should satisfy a single-parameter scaling
law
\begin{equation}
\rho_c(T) \propto \alpha_c g\left(\frac{T}{\Delta}\right),
\label{scaling0}
\end{equation}
where $g(x)$ is a scaling function, $\alpha_c$ is a
doping-dependent coefficient, and $\Delta$ is the maximal value of
the pseudogap. The detailed form of $g(x)$ is unknown. However, as
will be shown below, an approximate but accurate expression for
$g(x)$ can be obtained by simple theoretical analysis.

\begin{figure}[ht]
\includegraphics[width = 8cm]{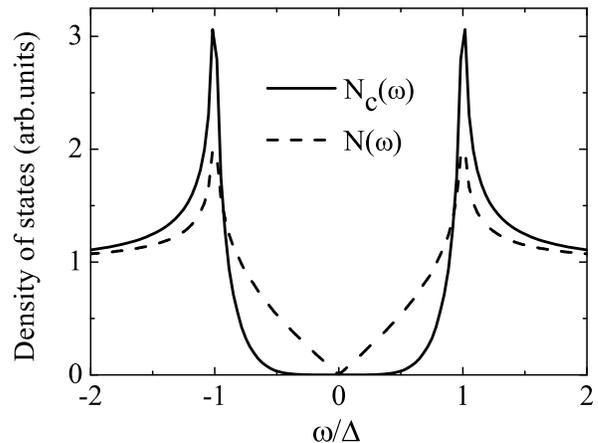}
\vskip-0.5cm\caption{Energy dependence of the effective c-axis
tunneling density of states $N_c(\omega)$. The density of states
of the d-wave gapped quasiparticles $N(\omega)$ is also shown for
comparison. $N(\omega)$ is defined by Eq. (\ref{nc}) with $t_c(k)$
= 1.\label{fig1}}
\end{figure}

$\rho_c$ is determined by the interlayer hopping integral and the
scattering mechanism. It can be obtained by evaluating the
current-current correlation function in the linear response
theory. In the highly anisotropic cuprate superconductors, the
c-axis hopping can be considered as perturbation. In this case,
the c-axis conductivity $\sigma_c$ is given by
\begin{equation}
\sigma_c(T) \propto -\sum_k \int d \, \omega v_c^2(k)
A^2(k,\omega) \frac{\partial f(\omega)}{\partial \omega},
\label{sigma0}
\end{equation}
where $v_c(k)\propto t_c(k)$ is the c-axis velocity of electrons
and $f(\omega)$ is the Fermi function. $A(k,\omega)$ is the
spectral function of pseudogapped electrons. At present, an
accurate microscopic description of $A(k,\omega)$ is not available
since the mechanism of pseudogap is still unclear. However,
phenomenologically one can assume $A(k,\omega)$ to have the form
\begin{equation}
A(k,\omega) = \frac{1}{\pi} \frac{\Gamma}{(\omega - E_k)^2 +
\Gamma^2}, \label{ak}
\end{equation}
where $E_k = \pm\sqrt{\varepsilon_k^2 + \Delta^2 \cos^2(2\phi_k)}$
is the energy dispersion of pseudogapped electrons, $\phi_k =
\arctan(k_y/k_x)$ and $\varepsilon_k$ is the usual normal state
energy measured from the Fermi level. $\Gamma$ is the linewidth,
proportional to the scattering rate. In the normal state above the
pseudogap phase, $\rho_{ab}$ varies linearly with T. It suggests
that $\Gamma$ scales linearly with T. This linear T dependent
$\Gamma$ is a basic assumption made in the marginal Fermi liquid
theory \cite{varma89}. It gives a natural account of many
experimental results, including the linear in-plane resistivity.
Here we will also adopt this assumption. Substituting Eq.
(\ref{ak}) into Eq. (\ref{sigma0}), it can be shown that to the
leading order approximation in $\Gamma$, $\sigma_c$ is given by
\begin{equation}
\sigma_c(T) \propto - \frac{1}{T} \int d\omega N_c(\omega)
\frac{\partial f(\omega)}{\partial \omega}, \label{sigma1}
\end{equation}
where $N_c(\omega)$ is an effective measure of the c-axis
tunneling probability of electrons
\begin{equation}
N_c(\omega) = \frac1V\sum_k t^2_c(k)\delta(\omega - E_k),
\label{nc}
\end{equation}
and $V$ is the system volume. $N_c(\omega)$ reduces to the density
of states $N(\omega)$ of d-wave gapped electrons if $t_c(k)$ in
Eq. (\ref{nc}) is set to 1. Figure \ref{fig1} compares the energy
dependence of $N_c(\omega)$ with $N(\omega)$.
It shows clearly that the c-axis hopping of low-energy
quasiparticles is strongly suppressed by $t_c(k)$ and the low
energy $N_c(\omega)$ is vanishingly small compared with that at
$\omega \sim \Delta$. Thus if the temperature considered is not
too low compared with $\Delta$, the contribution of low energy
electrons to the c-axis hopping can be ignored. This has motivated
us to replace approximately $N_c(\omega)$ by a step function,
\begin{equation}
N_c(\omega) \sim N_0 \theta(\omega - \Delta), \label{nc-approx}
\end{equation}
where $N_0$ is an average value of $N_c(\omega)$ at $\omega
\gtrsim \Delta$. Here $\Delta$ should be slightly smaller than the
true maximum pseudogap since the contribution from the states with
$\omega < \Delta$ should be effectively included in the above
approximation. Apparently this is a crude approximation, but it
allows us to get an analytical expression for $\sigma_c$. By
substituting (\ref{nc-approx}) into (\ref{sigma1}), we find that
$\rho_c$ is approximately given by
\begin{equation}
\rho_c(T) = \sigma_c^{-1}(T) \approx \frac{\alpha_c T}{\Delta}
\exp\left(\frac{\Delta}{T}\right) \label{rhoc}
\end{equation}
in the limit $T_c \ll T \ll \Delta$. We have tested this formula
with numerical calculations without taking the step-function
approximation for $N_c(\omega)$. We find that Eq. (\ref{rhoc})
does fit the numerical curves very well in the above limit. Though
this formula is derived within a limited temperature regime, it
captures qualitatively the main features of $\rho_c$ in the whole
temperature range in the normal state. In the low temperature
limit $T \ll \Delta$, Eq. (\ref{rhoc}) is thermally activated. In
the high temperature limit $T \gg \Delta$, $\rho_c$ varies
linearly with T. Both agree with the experimental observations.

\section{Comparison with experiments}

To test our single-parameter scaling conjecture, we have analyzed
the experimental data of $\rho_c$ published for
Bi$_2$Sr$_2$CaCu$_2$O$_{8+\delta}$ (Bi2212) by Watanabe et al.,
\cite{watanabe97, watanabe00} by Chen et al., \cite{chen98} and by
Guira et al., \cite{sarti03} for
Bi$_2$Sr$_2$Ca$_2$Cu$_3$O$_{10+\delta}$ (Bi2223) by Fujii et al.,
\cite{fujii02} and for YBa$_2$Cu$_3$O$_{7-\delta}$ (Y123) by Yan
et al. \cite{yan95} and by Babic et al.. \cite{babic99} We find
that all these experimental data of $\rho_c$ can be well fitted by
Eq. (\ref{rhoc}) nearly in the entire temperature range. Fig.
\ref{fig2}(a)-\ref{fig2}(c) show the scaling behavior of $\rho_c$
for Bi2212, Bi2223, and Y123, respectively. For Bi2212, we only
shoed the data published by Watanabe et al. \cite{watanabe97} for
clarity. The experimental data of Bi2212 by the same group
\cite{watanabe00} can be also scaled onto the same curve. The
experimental data of Bi2212 published by other two groups
\cite{chen98, sarti03} are showed in Fig. \ref{fig3}. The fitting
parameters $\Delta$ and $\alpha_c$ for Bi2212, Bi2223, and Y123 at
different doping levels are given in Table I-V. The
doping-dependence of $\Delta$ and comparison with the ARPES data
will be discussed later.
\begin{widetext}
\begin{center}
\begin{figure}[ht]
\includegraphics[width = 18cm]{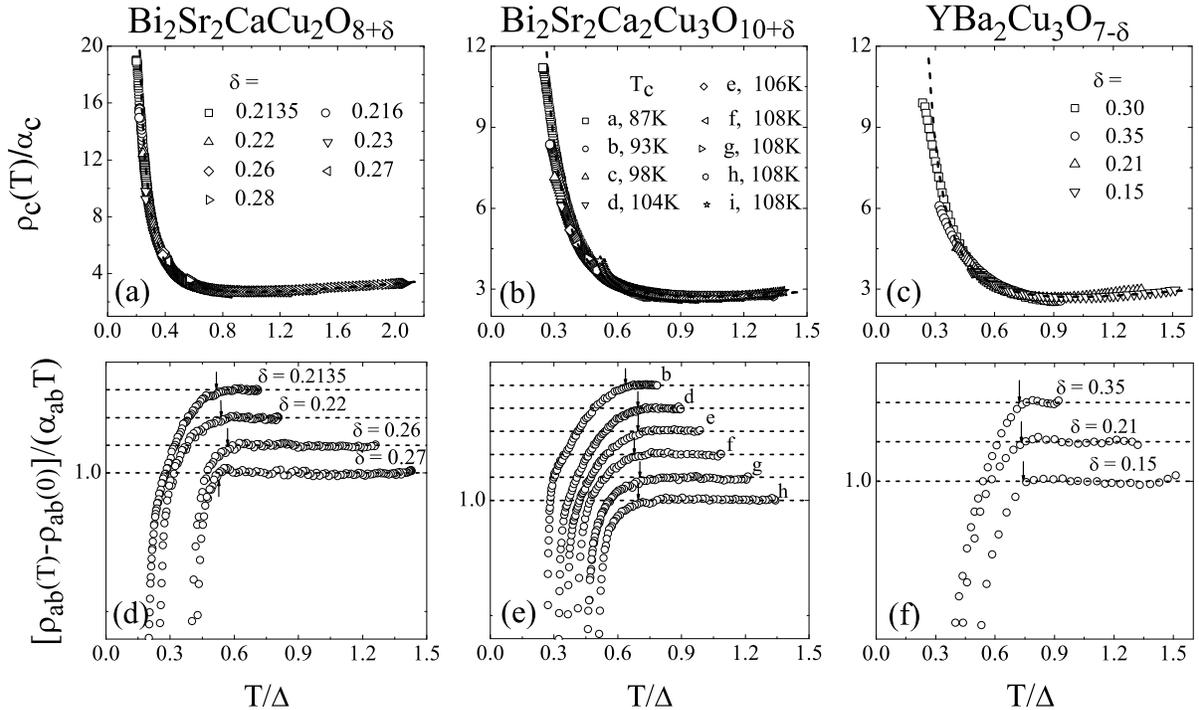}
\vskip-0.5cm\caption{Scaling behaviors of the c-axis and the
in-plane resistivities. (a)-(c), the scaling curves of
$\rho_c/\alpha_c$ versus T/$\Delta$ for the experimental data
published by Watanabe et al. \cite{watanabe97} for Bi2212, by
Fujii et al. \cite{fujii02} for Bi2223, and by Yan et al.
\cite{yan95} ($\delta = 0.3$) and by Babic et al. \cite{babic99}
for Y123, respectively. The parameters $\alpha_c$ and $\Delta$ are
determined by fitting the experimental data with Eq. (\ref{rhoc})
well above T$_c$. The dashed lines in (a)-(c) denote the scaling
function $g(x) = x \exp(1/x)$. (d)-(f), the corresponding
normalized in-plane resistivities [$\rho_{ab}(T)-
\rho_{ab}(0)$]/($\alpha_{ab}$T) as a function of T/$\Delta$ is
shown for Bi2212, Bi2223, and Y123, respectively. Different curves
are shifted vertically from each other for clarity.  Both
$\alpha_{ab}$ and $\rho_{ab}(0)$ can be determined by fitting the
experimental data in the linear regime of $\rho_{ab}$ with Eq.
(\ref{rhoab}). In this linear regime, the ratio $[\rho_{ab}(T)-
\rho_{ab}(0)]/(\alpha_{ab}T)$ is equal to $1$ within measurement
errors. However, below a doping dependent temperature T$^\ast$,
which is commonly defined as the onset temperature of the
pseudogap, $\rho_{ab}$ begins to deviate from this linear-T
behavior and $[\rho_{ab}(T)- \rho_{ab}(0)]/(\alpha_{ab}T)$  drops
quickly with decreasing temperature, as marked roughly by the
arrows in (d)-(f). \label{fig2}}
\end{figure}
\end{center}
\end{widetext}

The measurement data deviate slightly from the scaling curve near
T$_c$. This is due to the superconducting fluctuations
\cite{ioffe93}. This deviation is not unexpected since in this
case the single-parameter scaling law (2) should be modified to
include the contributions from the superconducting fluctuations.
The above analysis shows that the scaling behavior of $\rho_c$ is
universal and the scaling function $g(x)$ is approximately given
by
\begin{equation}
g(x) = x \, \exp\left(\frac1x\right) \label{scaling-func}
\end{equation}
independent of the doping concentration as well as the chemical
structures for these multilayer compounds. Thus there is only one
energy scale characterizing the c-axis dynamics in the pseudogap
phase. It suggests that the pseudogap is the only energy scale
governing the quasiparticle excitations at the antinodal points.

\begin{figure}[h]
\includegraphics[width = 8cm]{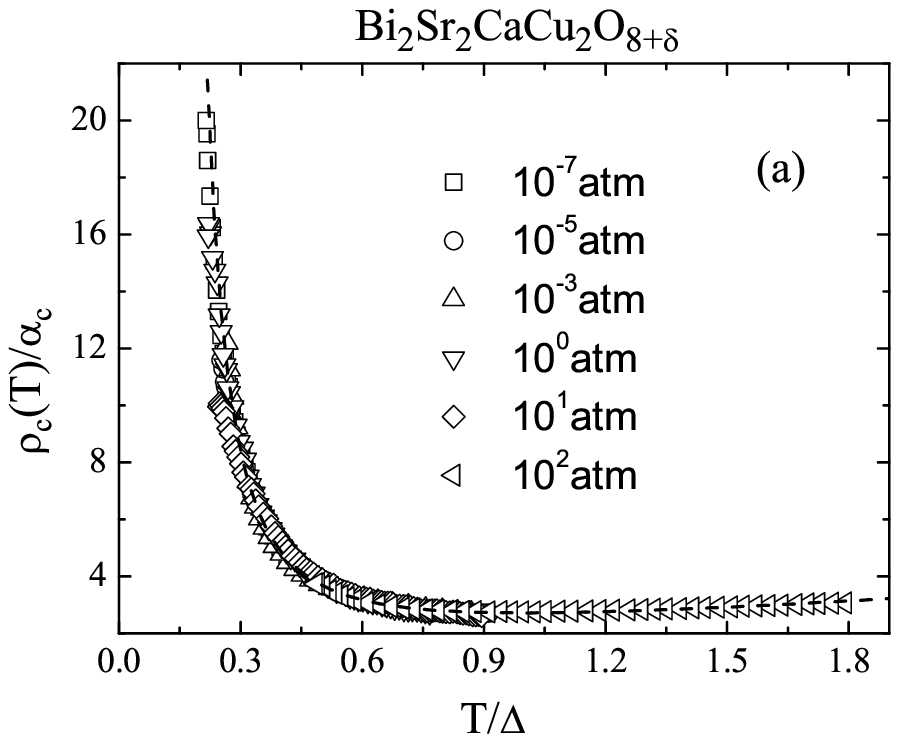}
\includegraphics[width = 8cm]{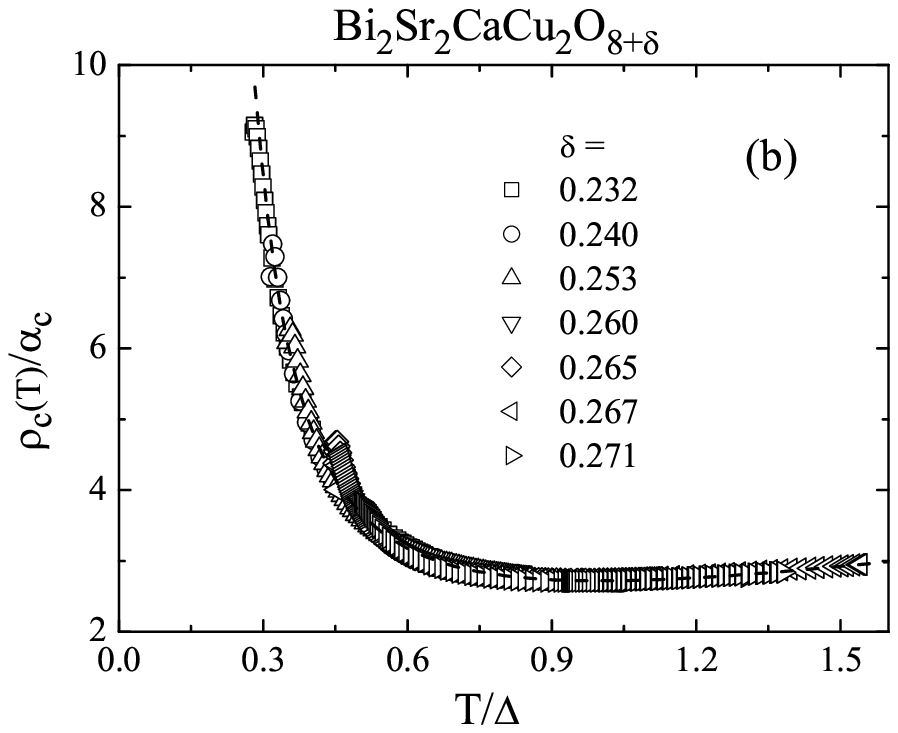}
\caption{The scaled c-axis resistivity for Bi2212 published by
Chen et al (a)[\onlinecite{chen98}] and by Sarti et al
(b)[\onlinecite{sarti03}]. \label{fig3}}
\end{figure}

\begin{table}[h]
\caption{The fitting parameters $\Delta$, $\alpha_c$,
$\alpha_{ab}$, and $\rho_{ab}(0)$ for Bi2212 published in Ref.
[\onlinecite{watanabe97}] (the first 7 rows) and Ref.
[\onlinecite{watanabe00}] (the rest 4 rows). The units of $
\alpha_{ab} $ and $\rho_{ab}(0)$ are $10^{-6} \Omega \mbox{cm}/K$
and $ 10^{-4}\Omega \mbox{cm} $, respectively. \label{bi2212}}
\begin{ruledtabular}
\begin{tabular}{cccccc}
$\delta$ & T$_c$ (K) & $\Delta$ (K) & $\alpha_c (\Omega \mbox{cm})$ & $\alpha_{ab}$ & $\rho_{ab}(0)$ \\
\hline
0.2135 & 71 & 416.6 & 2.74 & 2.60 & 1.3 \\
0.216 & 77 & 397.4 & 2.56  & - & - \\
0.22 & 83 & 367.3 & 2.31  & 1.78 & 1.1 \\
0.23 & 86 & 344.5 & 1.95  & - & - \\
0.26 & 88 & 232.9 & 1.07  & 1.03 & 0.22 \\
0.27 & 85 & 207.1 & 0.82 & 0.78 & 0.18 \\
0.28 & 79 & 143.3 & 0.56  & - & - \\
\\
0.25 & 90 & 255 & 1.14 & - & - \\
0.26 & 86 & 217 & 0.91  & - & - \\
0.27 & 83 & 184.7 & 0.67 & - & - \\
0.28 & 79 & 150.1 & 0.34  & - & -
\end{tabular}
\end{ruledtabular}
\end{table}

\begin{table}[ht]
\caption{The fitting parameters $\Delta$ and $\alpha_c$ for Bi2212
published by Chen et al. [\onlinecite{chen98}]. \label{bi2212a}}
\begin{ruledtabular}
\begin{tabular}{cccc}
Pressure(atm) & T$_c$(K) & $\Delta$ (K) & $\alpha_c (\Omega
\mbox{cm})$ \\
\hline
$10^{-7}$ & 76 & 373.4 & 3.58 \\
$10^{-5}$ & 81 & 329.4 & 3.44 \\
$10^{-3}$ & 89 & 343.5 & 2.55 \\
$10^{0}$  & 85 & 394.2 & 1.37 \\
$10^{1}$  & 80 & 334.8 & 1.01 \\
$10^{3}$  & 78 & 164.1 & 0.59
\end{tabular}
\end{ruledtabular}
\end{table}

\begin{table}[ht]
\caption{The fitting parameters $\Delta$ and $\alpha_c$ for Bi2212
published by Giura et al. [\onlinecite{sarti03}]. \label{bi2212b}}
\begin{ruledtabular}
\begin{tabular}{cccc}
$\delta$ & T$_c$ (K) & $\Delta$ (K) & $\alpha_c (\Omega \mbox{cm})$ \\
\hline
0.232 & 87.4 & 321.89 & 1.99 \\
0.240 & 88.1 & 281.62 & 1.64 \\
0.253 & 92.0 & 263.84 & 1.17 \\
0.260 & 87.0 & 194.8 &  0.69 \\
0.265 & 85.9 & 192.96 & 0.54 \\
0.267 & 86.2 & 191.69 & 0.34 \\
0.271 & 87.6 & 177.88 & 0.30
\end{tabular}
\end{ruledtabular}
\end{table}

\begin{table}[h]
\caption{The fitting parameters $\Delta, \alpha_c, \alpha_{ab}$,
and $\rho_{ab}(0)$ for Bi2223 published in Ref.
[\onlinecite{fujii02}]. The units of $ \alpha_{ab} $ and
$\rho_{ab}(0)$ are $10^{-6} \Omega \mbox{cm}/K$ and $
10^{-5}\Omega \mbox{cm} $, respectively. \label{bi2223}}
\begin{ruledtabular}
\begin{tabular}{cccccc}
Samples & T$_c$ (K)& $\Delta$ (K) & $\alpha_c (\Omega \mbox{cm})$ & $\alpha_{ab}$ & $\rho_{ab}(0)$ \\
\hline
a & 87 & 403.29 & 7.52 & - & - \\
b & 93 & 378.74 & 6.18  & 2.34 & -2.21 \\
c & 98 & 361.74 & 5.38 & - & - \\
d & 104 & 330.17 & 4.15  & 2.21 & -4.45 \\
e & 106 & 300.65 & 3.4 & 2.06 & -5.24 \\
f & 108 & 272.72 & 2.78  & 1.85 & -5.81 \\
g & 108 & 245.46 & 2.28 & 1.77 & -6.02 \\
h & 108 & 221.42 & 1.85  & 1.57 & -4.77 \\
i & 108 & 212.85 & 1.03  & - & -
\end{tabular}
\end{ruledtabular}
\end{table}

\begin{table}[h]
\caption{ The fitting parameters $\Delta, \alpha_c, \alpha_{ab}$,
and $\rho_{ab}(0)$ for Y123 published in Ref. [\onlinecite{yan95}]
($\delta = 0.3$) and Ref. [\onlinecite{babic99}](rest).
\label{y123}}
\begin{ruledtabular}
\begin{tabular}{cccccc}
$\delta$ & T$_c$ (K) & $\Delta$ (K) & $\alpha_c (m\Omega
\mbox{cm})$ & $\alpha_{ab}(\mu \Omega \mbox{cm}/K)$
& $\rho_{ab}(0)(\mu \Omega \mbox{cm})$ \\
\hline
0.35 & 61.4 & 317.49 & 10.41 & 1.36 & 1.45 \\
0.21 & 83.7 & 221.87 & 3.83 & 0.86 & 4.77 \\
0.15 & 91.0 & 194.25 & 2.48 & 0.84 & 4.84 \\
0.30 & 65.0 & 318.21 & 105.01 & - & -
\end{tabular}
\end{ruledtabular}
\end{table}
It is interesting to compare the above result with the temperature
dependence of $\rho_{ab}$. In high temperatures, $\rho_{ab}$ shows
linear temperature dependence
\begin{equation}
\rho_{ab}(T) = \alpha_{ab} T + \rho_{ab}(0), \label{rhoab}
\end{equation}
where $\alpha_{ab}$ is the slope of the linear resistivity and
$\rho_{ab}(0)$ is the zero temperature resistivity extrapolated
from high temperature data of $\rho_{ab}$. Fig. \ref{fig2}(d)-(f)
show the experimental data of [$\rho_{ab}(T)-
\rho_{ab}(0)$]/($\alpha_{ab}$T) as a function of T/$\Delta$ for
Bi2212, Bi2223, and Y123, respectively. The slope $\alpha_{ab}$
and $\rho_{ab}(0)$ for Bi2212, Bi2223, and Y123 are given in Table
I, IV, and V, respectively. Here $\Delta$ is the pseudogap value
determined from $\rho_c$. An intriguing result revealed by Fig.
\ref{fig2}(d)-(f) is that T$^\star$ (arrows in the figure) is
approximately proportional to $\Delta$, i.e., $T^\star \propto
\Delta$. This result is reminiscent of the relationship between
the superconducting transition temperature T$_c$ and the
superconducting energy gap $\Delta_0$. For a BCS mean-field d-wave
superconductor, T$_c$/$\Delta_0 \sim 0.47$. Here the ratio
T$^\star/\Delta$ is about 0.55 for Bi2212, 0.66 for Bi2223, and
0.71 for Y123. These values are larger than the BCS value. This is
not unexpected since, as pointed out above, $\Delta$ obtained by
Eq. (\ref{rhoc}) is smaller than the true pseudogap.

\begin{figure}[ht]
\includegraphics[width = 8cm]{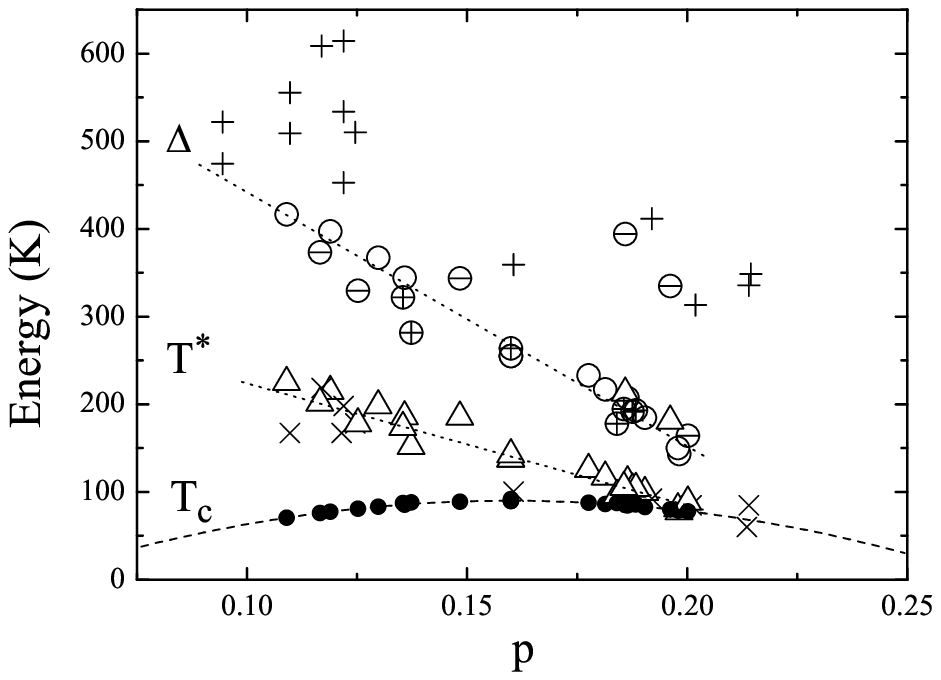}
\caption{Doping dependence of $\Delta$ ($\bigcirc$: Refs.
\onlinecite{watanabe97, watanabe00}; $\ominus$: Ref.
\onlinecite{chen98}; $\oplus$ Ref. \onlinecite{sarti03}),
T$^\star$ ($\triangle$: Refs. \onlinecite{watanabe97, watanabe00,
chen98, sarti03}), and T$_c$ ($\bullet$: Refs.
\onlinecite{watanabe97, watanabe00, chen98, sarti03}) for Bi2212.
T$^\star$ is extracted directly from the experimental data for the
four samples shown in Fig. \ref{fig2}(d). For all other samples,
T$^\star$ is determined from $\Delta$ by assuming T$^\star/\Delta
\sim 0.55$. The doping concentration $p$ is determined from T$_c$
using the empirical formula $T_c/T_{c,max} = 1- 82.6 (p - 0.16)^2$
(dashed line), T$_{c,max}$ is the maximum superconducting
transition temperature. For comparison, the ARPES measurement data
\cite{camp04} of $\Delta$ (+) and T$^\star$ ($\times$) are also
shown. The dotted lines are guide to eyes.\label{fig4}}
\end{figure}

Figure \ref{fig4} shows the doping-dependence of $\Delta$,
T$^\star$, and T$_c$ for Bi2212. For comparison, the ARPES data
published by Campuzano et al. \cite{camp04} for $\Delta$ and
T$^\star$ are also shown in the figure. The data of T$^\star$
obtained agree well with the ARPES data. Our results of $\Delta$
give a lower bound for the maximum pseudogap as expected. They are
less fluctuating than the ARPES data. It suggests that $\rho_c$ is
a good probe for the pseudogap.

The above comparison shows that $\rho_c$ satisfies a simple
single-parameter scaling law. We believe that the scaling function
$g(x)$ is universal and given approximately by Eq.
(\ref{scaling-func}) for all high-T$_c$ cuprates whose dominant
c-axis hopping integral between two neighboring CuO$_2$ layers is
coherent and has the in-plane momentum dependence given by Eq.
(\ref{tc}). This includes all multilayer compounds and some
single-layer compounds in which Cu atoms in the two neighboring
unit cells lie collinearly along the c-axis, such as
HgBa$_2$CuO$_{4+\delta}$. However, for other single-layer
compounds, for example Bi$_2$Sr$_2$CuO$_{6+\delta}$ and
La$_{2-x}$Sr$_x$CuO$_4$, Cu atoms of two adjacent CuO2 planes do
not lie collinearly along the c-axis. In this case, the c-axis
hopping integral has the form \cite{xiang98, marel00}
\begin{equation}
t_c(k) \propto \cos \frac{k_x}{2} \cos \frac{k_y}{2} \left(\cos
k_x - \cos k_y\right)^2.
\end{equation}
It vanishes along both nodal and antinodal directions. Therefore
the c-axis hopping is dominated by the quasiparticle excitations
in regions between the nodal and the antinodal points. This means
that the effective energy scale controlling the c-axis dynamics
should be the pseudogap at the momentum on the large Fermi surface
contour where $t_c(k)$ takes the maximum, rather than the
pseudogap itself. In this case, the scaling law Eq. (8) is still
valid in high temperatures. However, in low temperatures, Eq. (8)
needs to be modified since the contribution of nodal
quasiparticles to the c-axis transport is no longer negligible.

\section{Summary}

The interplay between the anisotropic c-axis hopping integral and
the pseudogap has profound consequence on the c-axis dynamics. It
leads to a universal scaling behavior for $\rho_c$ in the
pseudogap phase. The excellent agreement between our theoretical
analysis and experimental data suggests that the pseudogap is the
only energy scale governing the quasiparticle excitations around
the antinodal points in the normal state of high-T$_c$ materials.

Our analysis for the scaling behavior of $\rho_c$ is valid
generally, independent of the mechanism of normal state pseudogap.
We believe that it can be generalized and applied to other c-axis
transport quantities. Before closing the paper, we would like to
make a general conjecture: if $F_{c,e}$ is the electronic
contribution to a c-axis transport coefficient, then $F_{c,e}$
should satisfy a single-parameter scaling law
\begin{equation}
F_{c,e}(T) = \beta_c f_c\left(\frac{T}{\Delta}\right)
\end{equation}
at least at not too low temperatures in the pseudogap phase. Here
$\beta_c$ is a doping-dependent coefficient and $f_c(x)$ is a
corresponding scaling function. A thorough examination of this
conjecture, no matter being approved or disapproved, would shed
light on further understanding of properties of quasiparticle
excitations around the antinodal points. This will help us to
understand more about the mechanism of the high-T$_c$
superconductivity since the superconducting pairing is strongest
at the antinodal points.

\begin{acknowledgments}
We are grateful to T. Watanabe, X. H. Chen, and S. Sarti for
kindly sending us their experimental data published in Refs.
\onlinecite{watanabe97, watanabe00, fujii02, chen98, sarti03},
respectively. We thank L. Yu for valuable discussions. The work
was supported by the National Natural Sciences Foundation of
China.
\end{acknowledgments}

\end{document}